\documentclass[prb,twocolumn,showpacs,a4paper]{revtex4}
\usepackage{amssymb}
\usepackage{amsmath}
\usepackage{epsfig}
\usepackage{epstopdf}
\usepackage{bbm}
\usepackage{graphicx}
\usepackage{enumerate}
\usepackage{hyperref}
\usepackage{color}

\begin{document}

\title{Quantum criticality and non-Fermi-liquid behavior in a two-level two-lead quantum dot}

\author{Xin Wang and Andrew J. Millis
\\\vspace{.1in}\small{Department of Physics, Columbia University, New York, NY
10027}}

%\date{\today}

\begin{abstract}
Analytical and continuous-time quantum Monte Carlo methods are used to investigate the possibility of occupation switching and quantum criticality in a model of two quantum impurities coupled to two leads. A general discussion of potential occupancy-switching related quantum critical points is given, and a detailed analysis is made of a specific model which has been recently discussed. For spinless electrons, no phase transition is found. For electrons with spin, a critical value of the interaction strength separates a weak coupling regime in which all properties vary smoothly with parameters from a strong coupling phase in which occupation numbers vary discontinuously as level energies are changed. The discontinuity point is characterized by non-Fermi-liquid behavior. Results for self-energies and correlation functions are given. Phase diagrams are presented. \end{abstract}

\pacs{73.21.La, 71.10.Hf, 73.23.-b, 71.27.+a}

\maketitle

\section{Introduction}

Quantum impurity models (finite interacting systems coupled to infinite non-interacting reservoirs) give rise to interesting quantum critical behavior including the localization-delocalization transition of the macroscopic quantum tunneling problem,\cite{Leggett87} the non-Fermi-liquid physics associated with the under- and overscreened multichannel Kondo problem,\cite{Jayaprakash81,Fabrizio95} and the criticality related to the two-impurity Kondo model.\cite{Jones89,Jones89b, Affleck92}  Recently, a different model apparently exhibiting quantum criticality was uncovered in the context of experiments involving a quantum dot inserted into one arm of a Aharonov-Bohm interferometer.\cite{Yacoby95, Schuster97, Avinun05} As a gate voltage was varied between two Coulomb blockade peaks, the transmission phase was found to display a sharp ``phase lapse'' of $-\pi$. It is known that the phase lapse is associated with zeroes of the transmission amplitude or conductance,\cite{Oreg97, Yeyati00} and that many-body effects can play an important role in producing phase lapses.\cite{Hecht09}
One possible mechanism for  the vanishing of transmission amplitude and thus of  ``phase lapses'' is  ``population switching'' of different levels on the quantum dot.\cite{Silvestrov00, Konig05, Sindel05, Meden06, Kash07, Goldstein09a, Mueller09} This motivates an inquiry into the possibility of obtaining an abrupt population switching in models of multilevel dots coupled to leads. 

To model the phase lapse system, Golosov and Gefen\cite{Golosov06} introduced a two state quantum dot model, which they solved via mean-field theory in the spinless fermion approximation. They found population switching which led  to phase lapse behavior similar to that observed experimentally but raised the question whether the transition survives beyond the mean-field approximation. 
Meden and collaborators used numerical renormalization group methods to show that it in fact does not survive.\cite{Meden06,Karrasch07a,Karrasch07b}
In this paper we use analytical and numerically exact quantum Monte Carlo (QMC) techniques to go beyond mean-field theory and examine the switching transition. We find, in agreement with numerical renormalization group studies \cite{Sindel05, Meden06,Karrasch07a,Karrasch07b} of  that in the spinless case studied by Ref.~\onlinecite{Golosov06}, quantum fluctuations destroy the transition, whereas in the case of fermions with spin a population-switching  transition can exist at $T=0$. We present a brief discussion locating the transition in the general landscape of impurity-model quantum phase transitions. 

The rest of this paper is organized as follows. Section \ref{Model} presents the model that we study and a general discussion of the circumstances under which a quantum phase transition may take place. Section \ref{Analytics} outlines an approximate analytical approach to the problem which follows closely Hamann's analysis of the one-impurity Anderson model.\cite{Hamann69,Hamann70} Section \ref{Results} presents our numerical results and Section \ref{Conclusion} is a summary and conclusion.

\section{Model and methods \label{Model}}

\subsection{Overview}
In this section we present the models to be studied. While the initial impetus for the research comes from a specific realization of an interferometer involving a quantum dot \cite{Oreg97, Yeyati00} which implies an impurity model with a specific structure,\cite{Golosov06} it useful to present the results in a more general context.

A general quantum impurity model may be written:
\begin{equation}
H=H_{\rm dot}+H_{\rm lead}+H_{\rm mix},
\end{equation}
with
\begin{equation}
H_{\rm dot}=\sum_{ab}\varepsilon^{ab}
d^{\dagger}_ad_b+\sum_{a_1a_2b_1b_2}U^{a_1a_2b_1b_2}d^\dagger_{a_1}d^\dagger_{a_2}d_{b_1}d_{b_2}\label{hdot}
\end{equation}
describing the energetics of electrons in a set of states labeled by spin and orbital quantum numbers $a,b$,
\begin{equation}
H_{\rm lead}=\sum_{\lambda k}\varepsilon_k^\lambda
c^\dagger_{\lambda k}c_{\lambda k}\label{hlead}
\end{equation}
giving the energetics of electrons in a set of infinite leads labeled by a momentum (energy) quantum numbers $k$ and an index $\lambda$ which denotes lead and spin degrees of freedom, and
\begin{equation}
H_{\rm mix}=\sum_{k\lambda a}V_k^{a\lambda}c^\dagger_{\lambda
k}d_a+h.c.\label{hmix}
\end{equation}
giving the dot-lead hybridization. The effects of $H_{\rm lead}+H_{\rm mix}$ may be encoded in the hybridization function\cite{Werner2006a}
\begin{equation}
F^{ab}(-i\omega)=\sum_{k\lambda}\frac{V_k^{a\lambda}(V_k^{b\lambda})^*}{i\omega-\varepsilon_k^\lambda}.\label{hyb}
\end{equation}

We assume that the lead density of states and hybridization matrix elements are non-vanishing at the Fermi level where $\varepsilon^\lambda_k=0$ so that
\begin{equation}
F^{ab}(-i\omega_n\rightarrow 0)\rightarrow -i\Gamma^{ab}\mathrm{sgn}(\omega_n)+E^{ab}+{\cal O}\left(\frac{\omega_n}{E_0}\right)
\label{Flow}
\end{equation}
with $E_0$ a fixed energy characteristic of the conduction band band structure and hybridization.  We imagine integrating out energies on the scale of $E_0$ and higher and absorbing the resulting finite renormalizations into the parameters of $H$. We also incorporate the $E^{ab}$ into the dot parameters $\varepsilon^{ab}$ and choose the basis which diagonalizes the level broadening matrix  $\Gamma^{ab}$. Refs.~\onlinecite{Golosov06, Silva02} observed that transmission experiments which involve electrons injected and detected in particular combinations of leads may have phases which depend crucially on the form of the off-diagonal elements  when $F$ is written in the basis which diagonalizes the input and output currents, but these effects are not relevant for the considerations of this paper. 

We are interested in possible quantum critical behavior occurring as parameters in $H$ are varied. In all cases known to us, quantum critical behavior may be traced back to a crossing of eigenvalues of $H$ as some parameter is varied. Coupling to the leads may shift the position of the crossing (because $\varepsilon^{ab}\rightarrow \varepsilon^{ab}+E^{ab}$) and in addition may either promote the level crossing to a nontrivial critical point or convert it to a smooth crossover. For example, a singly occupied dot has a spin $S=1/2$ so that in the presence of a magnetic field one term in $H$ is ${\vec h}\cdot {\vec \sigma}$ with ${\vec \sigma}$ the usual Pauli matrices in spin space. If ${\vec h}$ is held parallel to a fixed direction (say ${\hat z}$) and the magnitude is varied from a positive to a negative value, then a level crossing occurs. In the generic case in which only one orbital is relevant and the coupling to the leads is antiferromagnetic the Kondo effect converts the level crossing into a smooth crossover. However if more than one orbital is relevant or if the Kondo coupling is ferromagnetic a form of multichannel or ferromagnetic Kondo criticality may occur.\cite{Jayaprakash81,Fabrizio95} One consequence of this criticality is a step in the spin polarization as the field is tuned  from a positive value through zero to a negative value. In a doubly occupied multiorbital dot, electronic singlet and triplet states are possible, and as a Hund's coupling parameter is varied through zero the singlet and triplet states may cross in energy. If two channels of conduction electrons are present, then coupling to the lattice will in the generic case convert the level crossing to a crossover, but in some situations the level crossing is shifted to a positive (antiferromagnetic) value of the Hund's coupling so that if appropriate symmetry conditions \cite{Jones89b,Affleck92} are imposed the crossing will become the ``Jones-Varma'' critical point.\cite{Jones89}

These examples  make it clear that quantum criticality is associated with the presence of a degeneracy point in the Hamiltonian of the isolated level and with a dependence of the state of the system on the direction from which the degeneracy point is approached. Further, one sees that  a high degree of symmetry must be enforced to obtain level crossing or quantum critical behavior, and that in particular the possible paths which the system can take through parameter space must be constrained. For example, in the Kondo example, if the field passes from $h=h_0{\hat z}$ to $h=-h_0{\hat z}$ by rotation, then the level crossing and any multichannel critical behavior are avoided. 

\subsection{Specific Models}
In this paper we consider two specific models. The first, studied by Golosov and Gefen \cite{Golosov06} is a two-level spinless fermion model. Adopting a pseudospin notation for the two orbitals of the impurity model we may write
\begin{equation}
H_{\rm dot}^{\rm spinless}={\vec \Delta}\cdot {\vec \tau}+U\hat{n}_1\hat{n}_2
\label{hdotspinless}
\end{equation}
with ${\vec \tau}$ the triplet of Pauli matrices acting on the orbital subspace and ${\vec \Delta}$ a generalized crystal field splitting. We also write for the level broadenings
\begin{equation}
\Gamma=\left(\begin{array}{cc}\Gamma^{11} & 0 \\0 & \Gamma^{22}\end{array}\right)
\label{Gammaspinless}
\end{equation}
The model is essentially the Anderson impurity model with a spin-dependent hybridization.  Degeneracy occurs in the one electron subspace as the effective crystal field splitting is tuned through zero. We will study the model assuming that the degeneracy point is approached along the direction ${\vec \Delta} \parallel {\hat z}$.  For more general directions of ${\vec \Delta}$ our numerical algorithm encounters a severe sign problem while the analytic theory becomes notationally much more complicated. However, the existence or not of a discontinuity as the system is tuned across the degeneracy point should not depend on how the point is approached.  

We also consider the same model, but for electrons with spin. A possible $H_{\rm dot}$ can be written as:
%\textcolor{red}{Important! Here the definition is equivalent to U11*2*n1up*n1do, in a difference with the calculations by a factor of 2! This must be fixed before publish}
\begin{equation}
\begin{split}
H_{\rm dot}^{\rm spin}&=\Delta\left(\hat{n}_{\rm 2,tot}-\hat{n}_{\rm 1,tot}\right)+U^{12}{\hat n}_{\rm 1,tot}{\hat n}_{\rm 2,tot}\\
&+U^{11}{\hat n}_{\rm 1,tot}(\hat{n}_{\rm 1,tot}-1)/2+U^{22}{\hat n}_{\rm 2,tot}({\hat n}_{\rm 2,tot}-1)/2\\
&+J_{\rm exch}{\vec S}_1\cdot{\vec S}_2+J_{\rm pair-hop}|0,2\rangle\langle2,0|
\end{split} 
\label{Hint1}
\end{equation}
which now has a richer level structure which depends on the dot occupancy. (Note that $\hat{n}_{a,{\rm tot}}=\sum_\sigma d^\dagger_{a\sigma}d_{a\sigma}$.) In the one or three  electron sector there are (counting spin and orbital degeneracy) four states. In the generic ${\vec \Delta} \neq 0$ case the orbital symmetry is lifted while time reversal symmetry would protect the spin degeneracy. If only one orbital is relevant one obtains the usual Kondo effect, but  in the special case ${\vec \Delta}=0$ a multichannel Kondo state may become possible.\cite{Fabrizio95} 

\begin{figure}[htb]
    \centering
    \includegraphics[width=6.5cm, angle=-90]{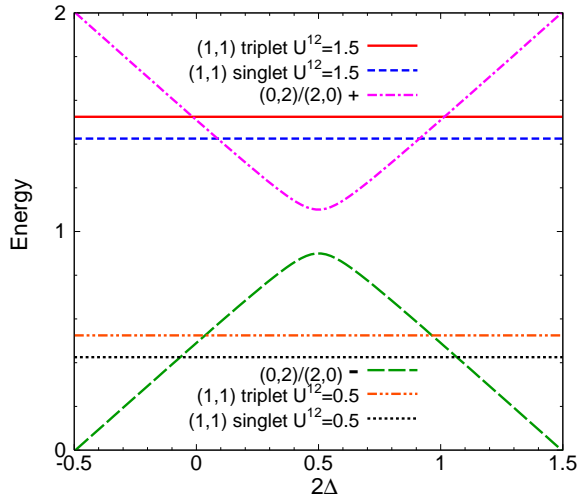}
    \caption{Two-electron energies calculated as functions of level splitting parameter $\Delta$ from $H_{\rm dot}$ with 
   $U^{11}=1.5$, $U^{22}=0.5$, 
   %(2 times the previous version because this U11 definition )
   $J_{\rm exch}=0.1$, $J_{\rm pair-hop}=0.1$ and $U^{12}=0.5$ and $1.5$.}
    \label{levelsketch}
\end{figure}

In the two electron sector there are the three members of a spin-1 triplet and three singlet states. In the occupation number basis we may label the singlet states as $|0,2\rangle$, $|1,1\rangle$ and $|2,0\rangle$. they are generically non-degenerate.  Fig.~\ref{levelsketch} shows possible evolutions of the two electron levels as the orbital splitting $\Delta$ is varied.  A positive Hund's coupling $J_{\rm exch}$ is assumed and two values of the intersite Coulomb interaction $U^{12}$ are shown. The three triplet states and the $|1,1\rangle$ singlet state have energy independent of $\Delta$.  For large $|\Delta|$ the lowest two electron state is one of $|0,2\rangle$ or $|2,0\rangle$. As $|\Delta|$ decreases, the energy of this favored state increases and level crossings occur. 

If the intersite Coulomb interaction is not too large [dotted line, double-dotted-dashed line (orange online)], there is a level crossing either to a triplet state (if $J_{\rm exch}<0$, not shown) or to the $|1,1\rangle$ singlet ($J_{\rm exch}>0$, shown). As noted above, the coupling to the leads would lead to a Kondo-quenching of the triplet states. A level crossing to this state would in general become a smooth crossover, but if particular symmetry conditions are satisfied a Jones-Varma fixed point would appear. We shall present arguments below indicating that the singlet-singlet crossing would become a smooth crossover. 

However, if the intersite Coulomb interaction is sufficiently large (short dashed blue line, solid red line) then the first level crossing will be between the $|0,2\rangle$ and $|2,0\rangle$-derived states. The degeneracy would be lifted by the ``pair-hopping'' terms, but in the absence of these terms a quantum critical point may ensue.

We observe from this analysis that a level crossing, and hence a quantum critical point, is expected only in the limit of vanishing pair-hopping. Physically, $J_{\rm exch}$ and $J_{\rm pair-hop}$ arise from inter-orbital exchange interactions. These interactions are likely to be very  small in the quantum dot case envisioned by Ref.~\onlinecite{Golosov06} where the orbitals correspond to spatially separated regions of the quanutm dot.

\subsection{Methods}
We shall be interested in the $d$ electron Green's function, defined on the Matsubara axis as:
\begin{equation}
G^{ab}(i\omega_n)=\int_0^\beta d\tau e^{-i\omega_n\tau}\langle
T_\tau d_a(\tau)d_b^{\dagger}(0)\rangle,
\end{equation}
$G^{ab}$ can be expressed in terms of the hybridization function and a self-energy resulting from the interaction term  as:
\begin{equation}
[G^{ab}(i\omega_n)]^{-1}=i\omega_n\delta_{ab}-\varepsilon^{ab}-F^{ab}(-i\omega_n)-\Sigma^{ab}(i\omega_n)
\end{equation}
and we shall mainly present results for $\Sigma$.

We will also present results for the density-density correlation function $W(\tau)$ defined as
\begin{equation}
W(\tau)=\left\langle T_\tau \left[{\hat n}_{\rm 1,tot}(\tau)-{\hat n}_{\rm 2,tot}(\tau)\right]
\left[{\hat n}_{\rm 1,tot}(0)-{\hat n}_{\rm 2,tot}(0)\right] \right\rangle\label{W}
\end{equation}

To study the models analytically we follow the techniques of Hamann,\cite{Hamann69,Hamann70} who wrote the single-impurity Anderson model\cite{Anderson61} as a functional integral, decouple the interaction with auxiliary fields, identify minima and consider the action associated with tunneling paths between minima. This maps the problem onto a macroscopic quantum mechanical model--one dimensional Coulomb gas\cite{Anderson69}--for which the key issue is the magnitude of the change in scattering phase shift between minima.

To study the models numerically we use a continuous-time QMC technique.\cite{Werner2006a, Werner2006b} Because we focus on a model without  complicated terms such as intradot hopping and exchange interactions, it is efficient to use the segment representation discussed in Ref.~\onlinecite{Werner2006b}. The method encounters a serious sign problem if an off-diagonal hybridization function is used, so we employ a basis in which $F$ is diagonal. In its current implementation the method requires that $H_{\rm dot}$ also be diagonal so we choose $\varepsilon^{ab}$ to be diagonal in the basis which diagonalizes $F$. As explained above, this does not affect our results.  Very recent work indicates that the computational cost of dealing with a non-diagonal $H_{\rm dot}$ need not be prohibitive,\cite{Lauchli09} so that in future work numerical studies of the case $\left[{\mathbf \varepsilon},\mathbf{F}\right]\neq 0$ may be feasible. 

We adopt a semicircular density of states for the conduction bands, 
\begin{equation}
\rho(\omega)=\frac{\sqrt{4t^2-\omega^2}}{2t^2},\ \ |\omega|<2t
\end{equation}
and choose ($k$-independent) hybridization parameters $V^{a\lambda}$ (Eq.~\eqref{hyb}) so that the level widths are much less than the bandwidth $4t$. The  frequency structure of the hybridization function affects only non-universal terms such as the quantitative location of the critical points. We also specialize to the case $U^{11}=U^{22}$ to simplify the presentation. 

Our calculations were performed on a parallel computer cluster with  20 dual core 2GHz nodes; a typical point requires up to about 15 hours of computer time on one CPU. The perturbation orders were typically $0\sim100$, but at the lowest temperatures orders up to $\sim 300$ were needed. The QMC technique is formulated at non-zero  temperature: in our work the lowest accessible temperature is $\beta t=800$ for calculations of Green's functions and self-energies and $\beta t=1600$ for a single point of the correlation function [Figs.~\ref{corrspinless}(b) and \ref{corrspinful}(b)].

\section{Analytical results \label{Analytics}}

The spinless fermion model may be viewed as one-orbital Anderson model with a spin-dependent hybridization. To study the model analytically we apply the methods of Hamann.\cite{Hamann69,Hamann70} Hamann writes the model as an imaginary-time path integral, re-expresses the interaction term as $\frac{U}{4}\left[(n_1+n_2)^2-(n_1-n_2)^2\right]$, decouples the $n_1+n_2$ and $n_1-n_2$ interactions with Hubbard-Stratonovich fields  $x$ and $\xi$ respectively and integrates out the fermions. In the situation considered by Hamann, the  Hubbard-Startonovich field  $x$ which couples to $n_1+n_2$ may be absorbed into the chemical potential and its fluctuations only provide finite renormalizations. In the situation of present interest, the field $x$   must be retained. After decoupling and integrating out the fermions the partition function becomes
\begin{equation}
Z=Z_0\int{\cal D}\xi{\cal D}x\ e^{S(\{\xi,x\})}
\label{pathint}
\end{equation}
with $Z_0$ the part of the partition function independent of $\xi,x$ and  (in the pseudospin notation of Eq.~\eqref{hdotspinless}
\begin{equation}
S={\rm Tr}\ \ln\left[ G_0^{-1}+\xi(\tau)\hat{\tau}_z+ix(\tau){\hat 1}\right]-\int_0^\beta d\tau\frac{\xi(\tau)^2+x(\tau)^2}{U}
\label{Sspinless}
\end{equation}
with the bare level Green function 
\begin{equation}
\begin{split}
G&_0(i\omega_n)=\\
&\left[(i\omega_n+\mu) {\hat 1} + \Delta\cdot\hat{\tau}_z +i\left({\bar \Gamma}{\hat  1}+\Gamma_z\hat{\tau}_z\right){\rm sgn}(\omega_n)\right]^{-1}
\end{split}
\label{G0spinless}
\end{equation}
and ${\bar \Gamma},\Gamma_z=\frac{1}{2}\left(\Gamma^{11}\pm\Gamma^{22}\right)$.

The analysis begins by solving the mean field equations $\partial S/\partial \xi=\partial S/\partial x=0$ for time independent $\xi,x$.  Defining $ix=\eta$, absorbing a Hartree shift of $\eta$ into the chemical potential  and performing the integrals in the wide bandwidth limit we obtain
\begin{eqnarray}
\begin{split}
\frac{2\pi\xi}{U}=&\arctan\frac{\Delta+{\bar \mu}+\xi+\eta}{\Gamma^{11}}\\&+\arctan\frac{\Delta-{\bar \mu}+\xi-\eta}{\Gamma^{22}},
\end{split}
\label{meanfieldspinlessxi}\\
\begin{split}
\frac{2\pi\eta}{U}=&-\arctan\frac{\Delta+{\bar \mu}+\xi+\eta}{\Gamma^{11}}\\&+\arctan\frac{\Delta-{\bar \mu}+\xi-\eta}{\Gamma^{22}},
\end{split}\label{meanfieldspinlesseta}
\end{eqnarray}
with ${\bar \mu}$ the  chemical potential shifted by the Hartree term from $\eta$.

\begin{figure}[htb]
    \centering
    \includegraphics[width=6.2cm, angle=-90]{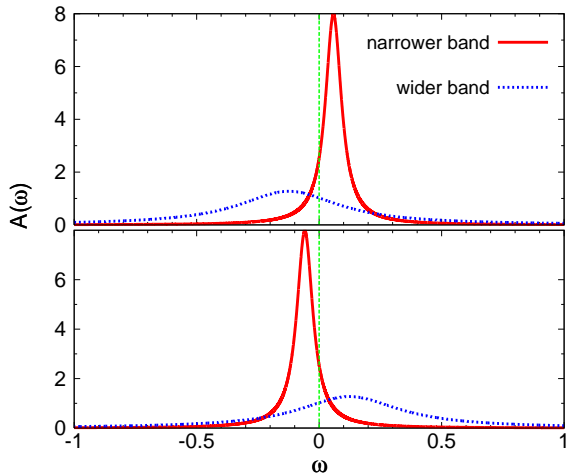}
    \caption{Electron spectral functions  $A(\omega)$ corresponding to Hartree-Fock solutions (Eqs~\eqref{meanfieldspinlessxi},\eqref{meanfieldspinlesseta}) for parameters $\Gamma^{11}=0.04$, $\Gamma^{22}=0.25$, $U=0.4$ and $\Delta=\mu=0$. The units of $\omega$ are defined by the values of $\Gamma^{11}$, $\Gamma^{22}$.}
    \label{meanfieldsolution}
\end{figure}

For small $U$ there is only one stable solution. For $U$ larger than a critical value there is a bifurcation and two stable solutions appear. The minimum $U$ occurs at the particle-hole symmetric point $\Delta={\bar \mu}=0$ where the transition is second order. For $\Delta$ or ${\bar \mu}\neq 0$ the transition shifts to higher $U$ and becomes discontinuous. At the particle-hole symmetric point linearizing the equations in $\eta,\xi$ shows that the critical $U$ is
\begin{equation}
U_c^{\rm ph}=\pi\sqrt{\Gamma^{11}\Gamma^{22}}
\label{Uc}
\end{equation}
while the eigenvector corresponding to the zero eigenvalue satisfies
\begin{equation}
\frac{\eta}{\xi}=-\frac{\sqrt{\Gamma^{22}}-\sqrt{\Gamma^{11}}}{\sqrt{\Gamma^{22}}+\sqrt{\Gamma^{11}}}
\label{eigenvalue}
\end{equation}
To interpret this result we note that $(\xi + \eta)$ gives the shift of level $1$  and  $(\xi - \eta)$ the  shift of level $2$. If $\Gamma^{22}\gg\Gamma^{11}$ then we see from Eq.~\eqref{eigenvalue} that the two mean field solutions correspond to a larger shift of the broad level and a smaller shift of the narrow level. However, what is important for the occupation is the level shift relative to the level width, and we see that $|\xi+\eta|/\Gamma^{11}\gg|\xi-\eta|/\Gamma^{22}$ so that the relative change in occupation of the narrow level is larger.  This structure  is found in the solution of the full nonlinear equations: in the limit of very different level widths the two solutions correspond to states in which the population of the narrow level changes substantially while that of the broader level shifts much less, with the total level occupancy differing in the two stable states.  An example is given in Fig.~\ref{meanfieldsolution}. We note, though that the important parameter is $\sqrt{\Gamma^{11}/\Gamma^{22}}$ so an extreme disparity in level widths is needed to produce a large difference in change in occupation.   The two solutions in general have different energy; for a fixed $\mu$ and $U$ there is a $\Delta$ ($=0$ in the particle-hole symmetric case) at which the energies cross.

Fluctuations of course alter the mean field predictions. In the strong coupling limit we follow Hamann and identify  the most important fluctuations as ``kinks'' in which the pair  $\xi,\eta$ tunnels from the neighborhood of one mean field solution to the neighborhood of the other. Hamann estimates the action associated with these tunnelling events  from the solution of a singular integral equation. The result is an expression for the partition function as a sum over kinks occurring at times $\tau_m$ with a logarithmic interaction between them
\begin{equation}
\begin{split}
&\frac{Z}{Z_0}\approx \sum_{n=0}^\infty f^{2n} \int_0^\beta\frac{d\tau_{2n}}{\tau_0}
\int_0^{\tau_{2n}-\tau_0}\frac{d\tau_{2n-1}}{\tau_0}\cdots
\int_0^{\tau_2-\tau_0}\frac{d\tau_{1}}{\tau_0}
\\
&\times \exp\left[{\bar \Delta}\sum_{i=1,3,..} (\tau_{i+1}-\tau_{i})+K\sum_{i\neq
j}(-1)^{i+j}\ln\left|\frac{\tau_i-\tau_j}{\tau_0}\right|\right]
\end{split}\label{Ztun}
\end{equation}
with $Z_0$ representing the non-tunnelling contributions to the partition function, $\tau_0$ a small cutoff time scale  of the order of  the larger of $1/\Gamma^{11,22}$, $f$ a tunnelling fugacity of order $\ln U\tau_0$ determined by the bare parameters in the problem, ${\bar \Delta}$ the difference of $\Delta$ from the critical value at which the two solutions are degenerate,  and, most importantly, an interaction coefficient $K$ determined by the change $\Delta \delta$ in scattering phase shifts between the different solutions:
\begin{equation}
K=\sum_{{\rm channels},a}\left(\frac{\Delta\delta_a}{\pi}\right)^2\approx\sum_{a}(\Delta n_a)^2
\label{Kspinless}
\end{equation} 
The second approximate equality applies in the wide bandwidth limit where the phase shift gives directly the level occupancy.

It is known \cite{Anderson69} that for $K<2$ the model  defined by Eq.~\eqref{Ztun} is in its screening phase: tunneling events proliferate and physical properties vary smoothly as parameters such as ${\bar \Delta}$ are varied, whereas for $K>2$ the model is in the unscreened or  localized phase in which tunnelling is suppressed and there is a discontinuous change as $\Delta$ is tuned across the critical value.

In the spinless case the maximal change in density in either channel for finite $U$ is $<1$ so the Hamann analysis indicates that the model is always in its screening phase and no phase transition is expected, in contradiction to the results of Ref.~\onlinecite{Golosov06}.  However, if spin is taken into consideration the number of channels is doubled and $K>2$ becomes possible.  In the model with spin  the various interactions require a multiplicity of Hubbard-Stratonovich fields.\cite{Sakai04} Solutions to the mean field equations involve non-zero values of the decoupling fields corresponding to the dominant interactions, with the other decoupling fields being unimportant.\cite{Lin08} Assuming that $U^{12}$ is dominant and neglecting for the moment the pair hopping term we write this contribution to the interaction term as $\frac{U^{12}}{4}\left(n_{\rm 1,tot}+n_{\rm 2,tot}\right)^2-\frac{U^{12}}{4}\left(n_{\rm 1,tot}-n_{\rm 2,tot}\right)^2$, decouple the two terms as before and obtain an action of the same form as Eq.~\eqref{Sspinless} except that all of the quantities are now to be interpreted as $4\times4$ matrices to include the spin degeneracy. The analysis proceeds as above except that an extra factor of two multiplies the $\arctan$ in  Eqs.~\eqref{meanfieldspinlessxi}, \eqref{meanfieldspinlesseta} as well as the parameter ${\bar \Delta}$ in Eq.~\eqref{Ztun}. The parameter $K$, now approximately $(\Delta n_{1\uparrow})^2+(\Delta n_{1\downarrow})^2+(\Delta n_{2\uparrow})^2+(\Delta n_{2\downarrow})^2$ may become larger than $2$, driving the system into the unscreened phase.  

In our work we choose $\Gamma^{11}/t=0.04$, $\Gamma^{22}/t=0.25$, implying $U_c=0.157...$ for the model with spin (at the particle-hole symmetric point).  Numerically solving the mean field equations we find that at $U\approx 0.32$ the change $\Delta n_{1\uparrow}\approx 0.85$ and $\Delta n_{2\uparrow}\approx 0.53$ so that the coupling constant exceeds $2$ and we would expect the model to become localized.

As observed above, the pair-hopping term, by mixing the two states $|0,2\rangle,|2,0\rangle$ will destroy the phase transition, leaving instead a smooth crossover.

Finally, we note that the $|1,1\rangle\rightarrow|2,0\rangle$ level crossing (shown for smaller $U^{12}$ in Fig.~\ref{levelsketch}), would involve an occupancy change of $1/2$ electron per spin,  in one orbital only, so that we would expect the corresponding model to be in its screened phase so that no phase transition would ensue.

\section{Numerical results \label{Results}}

\subsection{Spinless fermions}

\begin{figure}
    \centering
    \includegraphics[width=6cm, angle=-90]{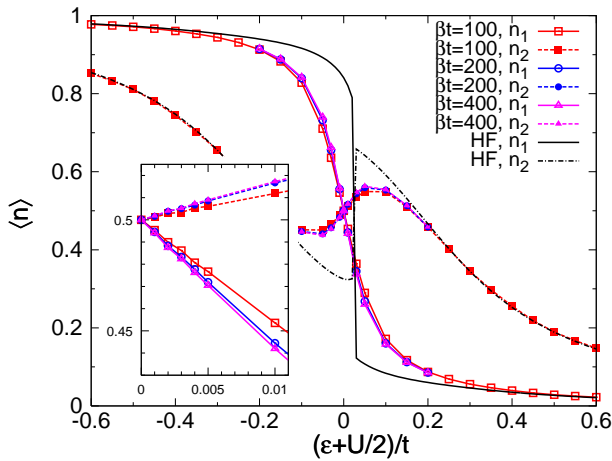}
    \caption{Main panel: occupancy of level 1 (solid curves) and level 2 (dashed curves) of spinless fermion model as function of mean level energy $\varepsilon=\varepsilon^{11}=\varepsilon^{22}$ shifted by $U/2$ obtained from Hartree Fock (HF, lines without symbols, black on-line) and numerically exact QMC calculations (lines with symbols, red, blue and magenta on-line).  HF is computed at $T=0$ while QMC is performed at different temperatures as shown in the legends. Inset: expanded view of energy dependence of dot occupancies in energy regime close to the particle-hole symmetric point. Parameters: $\Gamma^{11}/t=0.04$, $\Gamma^{22}/t=0.25$, $U/t=0.4$. }
\label{nvsmuspinless}
\end{figure}

Fig.~\ref{nvsmuspinless} shows our numerical results for the $d$-occupancy as a function of average dot energy in spinless case. We took the energy of two levels of the dot to be equal, i.e. $\varepsilon^{11}=\varepsilon^{22}=\varepsilon$. Level one is assumed to be narrower ($\Gamma^{11}/t=0.04$) and level two broader ($\Gamma^{22}/t=0.25$). The Coulomb interaction $U/t=0.4$.  The solid and dash-dotted lines present the Hartree-Fock results for the level $1$ and $2$ occupancies respectively. Within the Hartree-Fock approximation there is a first order phase transition at  $\varepsilon=\varepsilon^*\simeq0.03t -U/2$; as $\varepsilon$ is tuned through $\varepsilon^*$ the population of the two levels ``switches'' abruptly. However, the QMC results, indicated by lines with points in  Fig.~\ref{nvsmuspinless}, shows that at the value of $U$ studied here the phase transition is an artifact of the Hartree-Fock approximation. At half-filling point $\varepsilon=-U/2$, the only stable solution is $\langle n_1\rangle=\langle n_2\rangle=1/2$, while the weak temperature dependence (displayed in an expanded scale in the inset) indicates that the simulation has accessed the low temperature limit of the model.

\begin{figure}
    \centering
    \includegraphics[width=6.5cm, angle=-90]{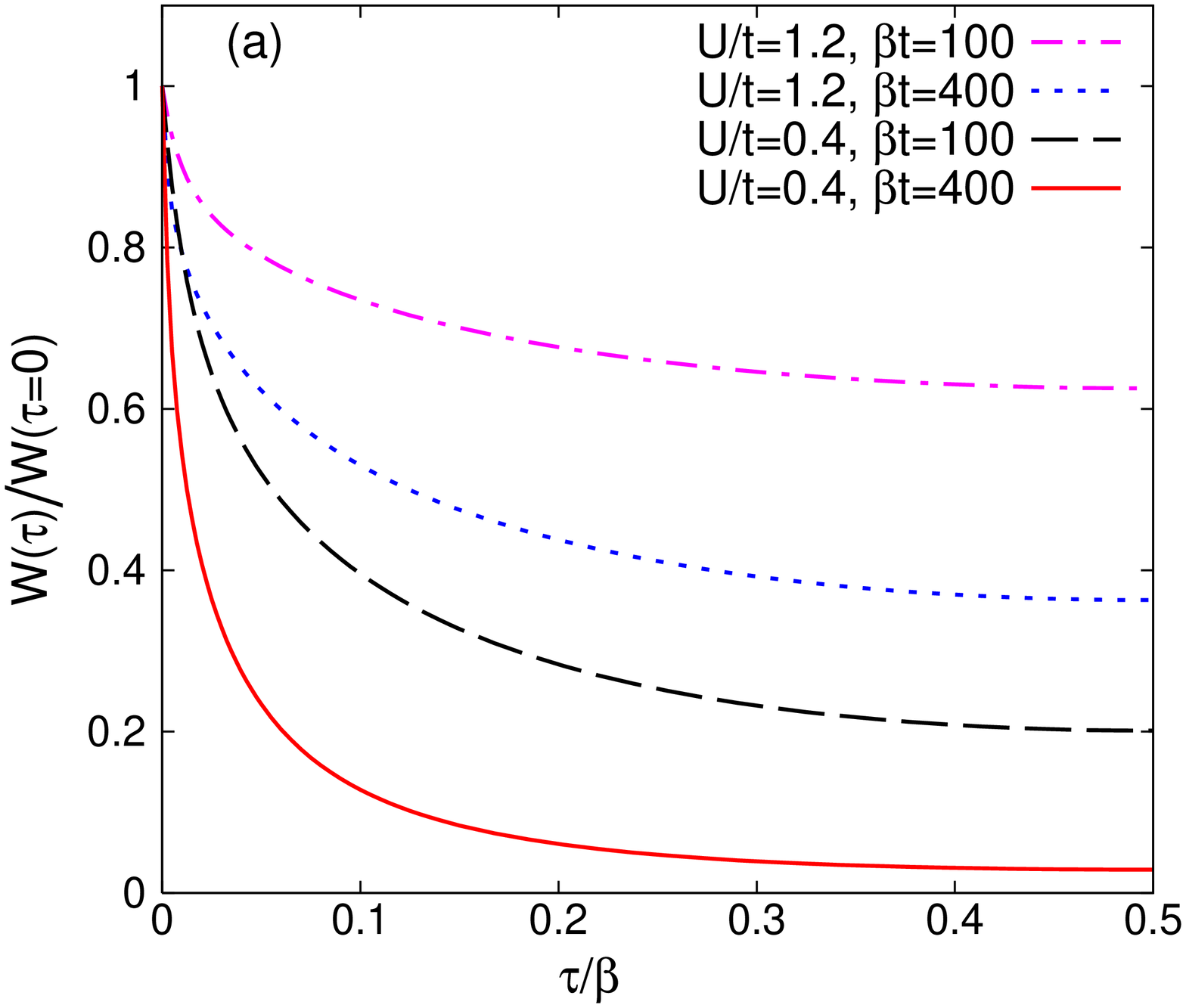}
    \includegraphics[width=6.5cm, angle=-90]{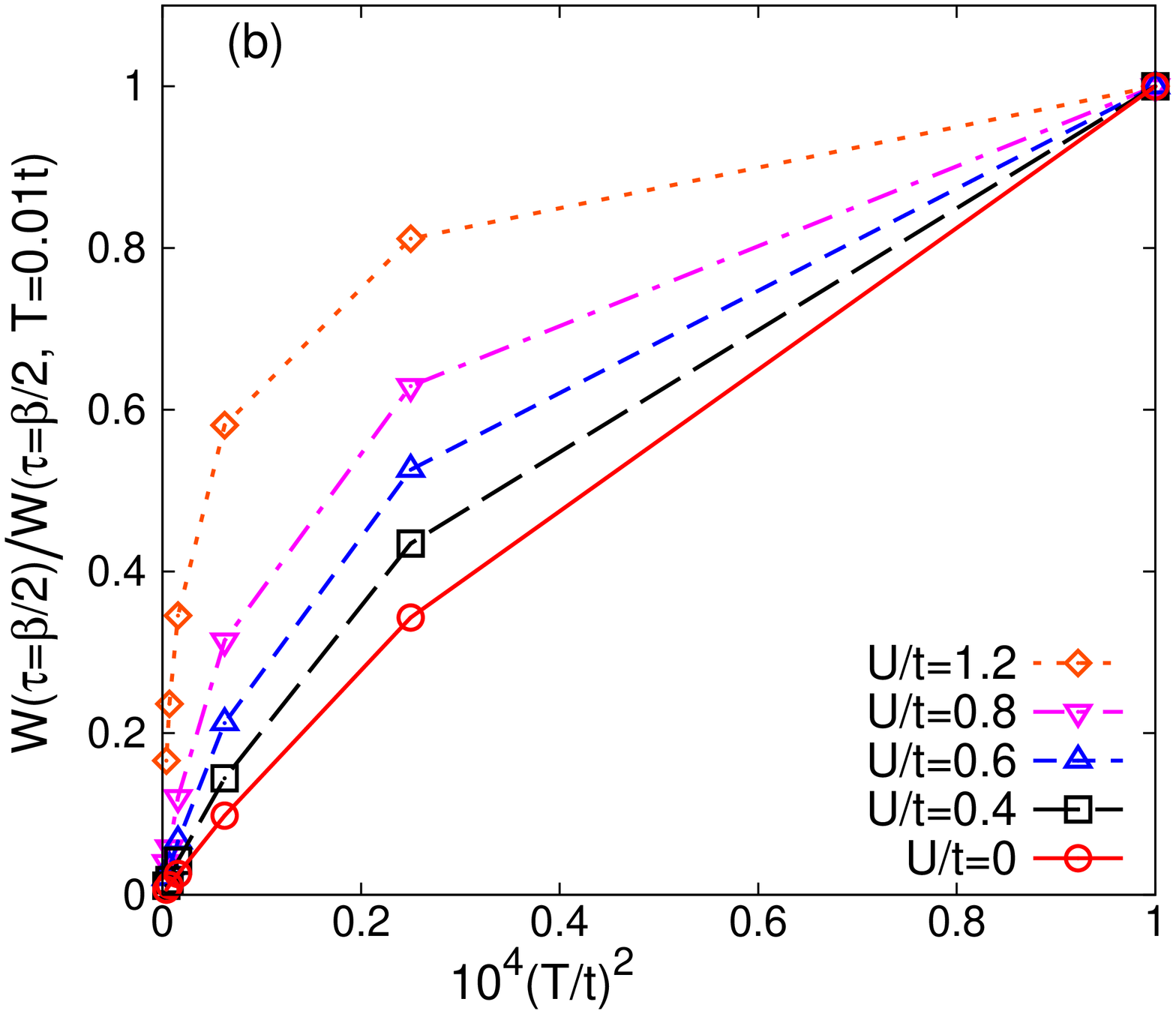}
    \caption{Imaginary time density-density correlation function $W$ (Eq.~\eqref{W}) computed at putative quantum critical point for spinless fermions at parameters indicated. (a) Full imaginary time dependence of $W(\tau)$ normalized to its $\tau=0$ value, indicating an obvious temperature dependence for both small and large $U$. (b) Correlation function evaluated at midpoint of imaginary time interval, normalized to value at $T=0.01t$, indicating $T^2$ dependence at low $T$. Parameters $\Gamma^{11}/t=0.04$, $\Gamma^{22}/t=0.25$.}
    \label{corrspinless}
\end{figure}

As further evidence of the absence of a phase transition in the spinless model we present in Fig.~\ref{corrspinless} the imaginary-time density-density correlation function (Eq.~\eqref{W}) computed at the particle-hole symmetric point for different interaction strengths at various temperatures. Panel (a) displays the full imaginary time dependence (note $W(\tau)=W(\beta-\tau)$). In both the intermediate ($U/t=0.4$) and the strong ($U/t=1.2$) interaction case the correlation function drops as $\tau$ is increased from $\tau=0$ and exhibits a minimum at $\tau=\beta/2$. The value $W(\tau=\beta/2)$ has a clear temperature dependence which is displayed in more detail in Fig.~\ref{corrspinless}(b): we see that all curves extrapolate to $0$. For $0\leq U/t\leq0.6$ our temperature range is sufficient to resolve clearly the $T^2$ behavior expected from Fermi liquid theory;  for $U/t=0.8, 1.2$ the temperatures numerically accessible to us are not low enough to establish a convincing $T^2$ dependence; the extrapolation to zero is evident. 

An analytical expression for the  Kondo temperature for the spinless model has been derived as:\cite{Haldane1978,Meden06,Kash07,Lee07}
\begin{equation}
T_K\approx0.2\frac{\sqrt{U(\Gamma^{11}+\Gamma^{22})}}{\pi}\exp\left[\frac{\pi\varepsilon(U+\varepsilon)}{2U(\Gamma^{11}-\Gamma^{22})}\ln\left(\frac{\Gamma^{11}}{\Gamma^{22}}\right)\right]
\end{equation}
where the prefactor 0.2 has been set by recovering the result of Anderson model\cite{Haldane1978} when taking $\Gamma^{11}=\Gamma^{22}$. For the $0.4\leq U/t\leq1.2$ studied in this section the Kondo temperature varies from $T_K/t=0.006$ for $U/t=0.4$ to $T_K/t=0.0006$ for $U/t=1.2$. Our lowest accessable temperature, $T/t=1/800$, is about twice the $U=1.2$ $T_K$, explaining why the first stages of the crossover to Fermi liquid behavior are evident even in this case. 

Finally, we consider  the electron self energy. At the particle-hole symmetric point the Matsubara axis self energy is purely imaginary and in a Fermi liquid state would vanish proportionally to $\omega_n$ as $\omega_n\rightarrow 0$.   To make the analysis more precise we note that  Fermi liquid theory implies (see e.g. Ref.~\onlinecite{Yamada70,Tsvelick83}) that  at low frequency and temperature the real-axis self-energy is approximately 
\begin{equation}
\Sigma(\omega)=(1-Z^{-1})\omega-i\frac{T_0}{2}A^2\left(\frac{\omega^2}{T_0^2}+\frac{\pi^2T^2}{T_0^2}\right)
\label{FLSE}
\end{equation} 
with $T_0$ a scale of the order of the Kondo temperature, $Z^{-1}\sim \Gamma/T_0$, and $A$ a number of the order of unity. This form implies that at the particle-hole symmetric point 
\begin{equation}
\Sigma(i\omega_n)=(1-Z^{-1})i\omega_n-i\frac{\pi^2}{2}A^2\frac{T^2}{T_0} {\rm sgn}(\omega_n)+...
\label{FLSEMat}
\end{equation}
where the ellipsis denotes terms of higher order in $\omega_n$. 

\begin{figure}
    \centering
    \includegraphics[width=6.5cm, angle=-90]{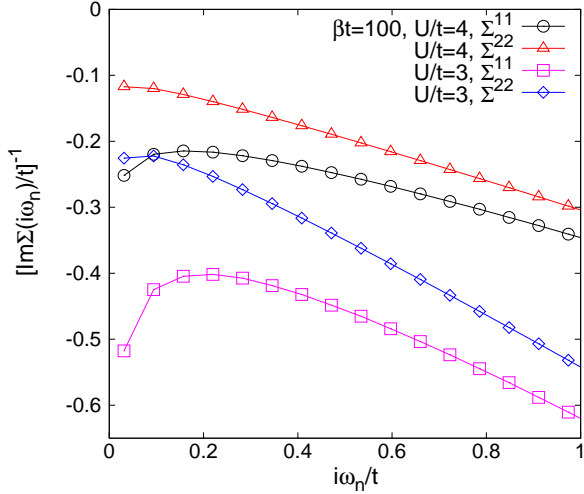}
    \caption{\emph{Reciprocal} of the imaginary part self energy as a function of Matsubara frequencies computed for spinless fermions at the  particle-hole symmetric point for  very large interaction $U/t=3,4$, with $\Gamma^{11}/t=0.04$, $\Gamma^{22}/t=0.25$, $\beta t=100$. The low frequency downturn is a indication of Kondo bahavior.}
    \label{seuscanspinless}
\end{figure}

Fig.~\ref{seuscanspinless} presents the imaginary part of the \emph{reciprocal} of the Matsubara-axis electron self energy $\Sigma(i\omega_n)$ computed for several temperatures at two very large $U$ values. The weak downturn of $\Sigma^{11}$ at $\omega_n/t\sim0.1$ is a signature of the crossover to Fermi-liquid behavior. Surprisingly, a hint of the crossover to the Fermi liquid behavior is evident even at temperatures much higher than the Kondo temperature. For $U/t=3$, the Kondo temperature $T_K/t=2\times10^{-6}$, so the beginning of the crossover (upturn in $1/{\rm Im}\Sigma)$ can be observed even for  $T\sim5000T_K$.

\subsection{Fermions with spin}

\begin{figure}
    \centering
    \includegraphics[width=6cm, angle=-90]{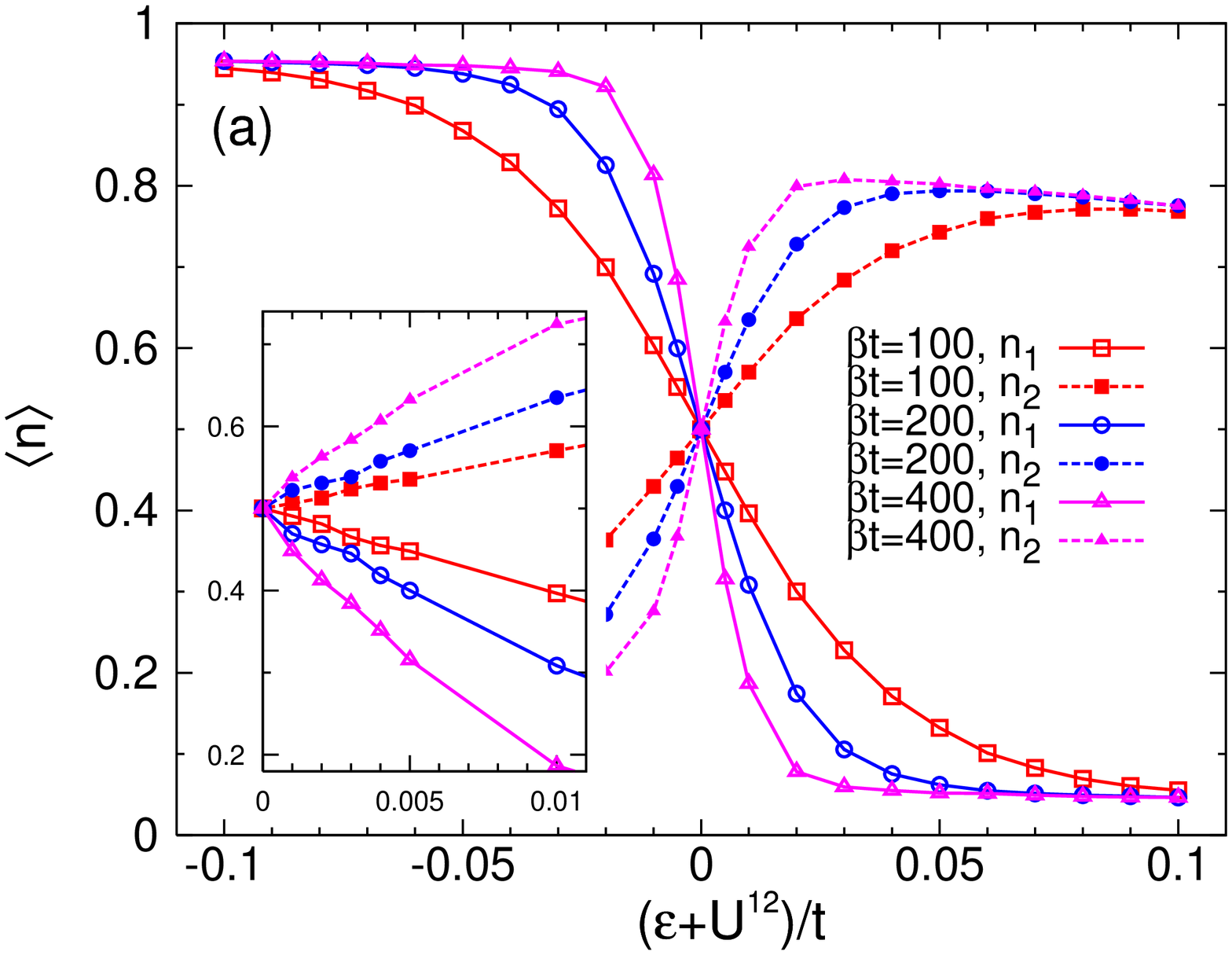}
    \includegraphics[width=6.2cm, angle=-90]{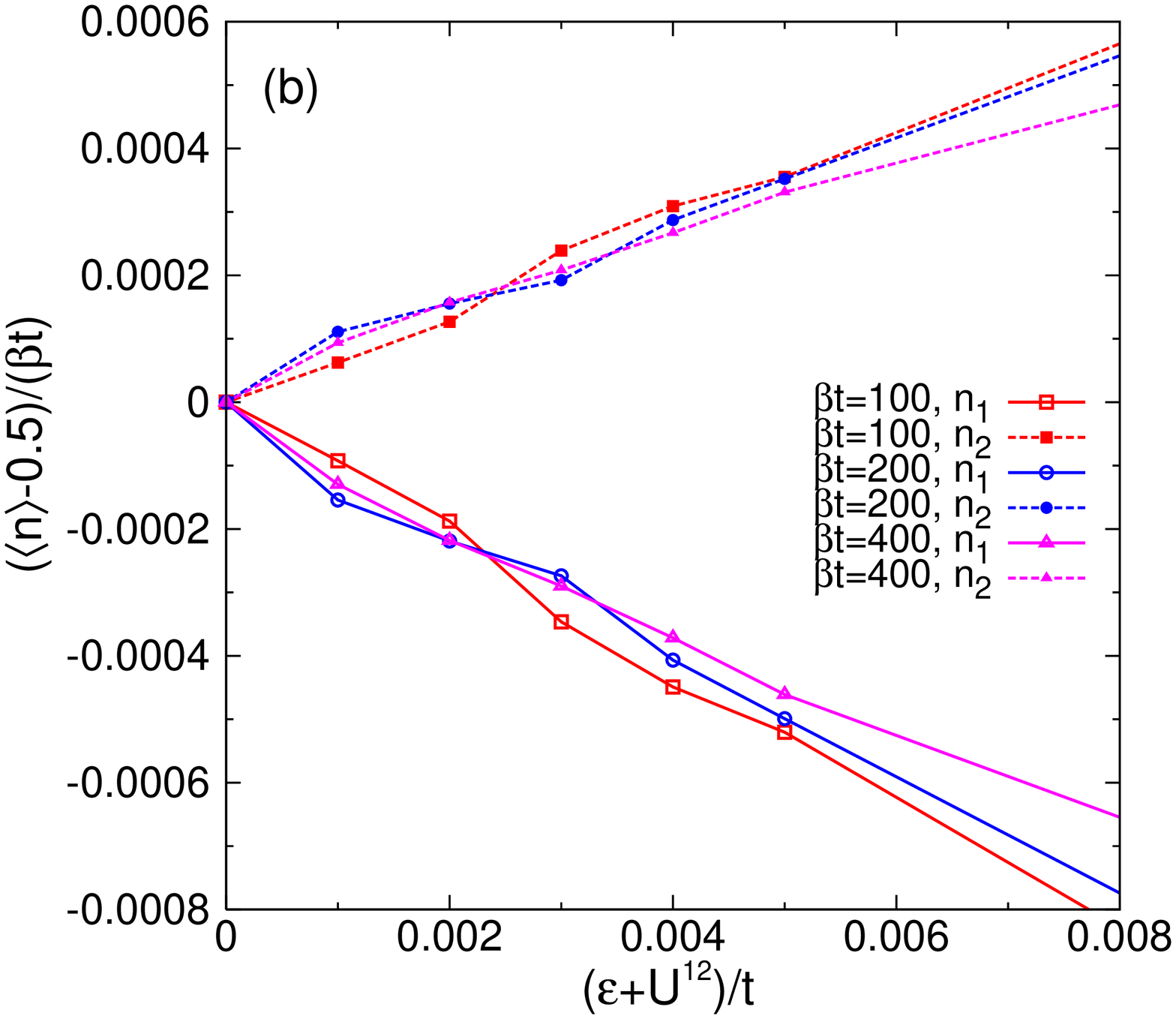}
    \caption{(a) Main panel: occupancy of level 1 (solid curves) and level 2 (dashed curves) for fermions with spin as function of mean level energy $\varepsilon=\varepsilon^{11}=\varepsilon^{22}$ shifted by $U^{12}$ such that the degeneracy point corresponds to $\varepsilon=0$ obtained from QMC calculations (red, blue and magenta online), at different temperatures as shown in the legends. Inset: expanded view of energy dependence of dot occupancies in energy regime close to the particle-hole symmetric point. (b) The same data plotted as $(\langle n\rangle-0.5)/(\beta t)$ displaying $1/T$ temperature dependence. Parameters: $\Gamma^{11}/t=0.04$, $\Gamma^{22}/t=0.25$, $U^{12}/t=0.4$. Note the x range is different from that of Fig.~\ref{nvsmuspinless}. }
    \label{nvsmuspinful}
\end{figure}

Fig.~\ref{nvsmuspinful}(a) shows our numerical results for the $d$-occupancy as a function of average dot energy for fermions with spin with $U^{11}=U^{22}=0$, the pair hopping neglected, the level energies and widths as in the spinless case, the moderate interaction $U^{12}=0.4t$ and various temperatures.  In contrast to the spinless case, a clear temperature dependence is evident. The panel (a) inset and panel (b) show that near the degeneracy point  $\langle n_1-n_2\rangle\sim1/T$, so that the dot occupancy exhibits the approximately Curie-Weiss behavior expected of a two state system with no mixing between the states, in sharp contrast to the spinless case.

At high $U^{12}$, the QMC calculation takes very long time to converge, especially at low temperatures.  We believe that the convergence difficulties are related to the strong suppression of tunnelling between  the two states. The expense of the calculations means that for the rest of the section  we restrict ourselves to the particle-hole symmetric points $\varepsilon=-U^{12}$, and enforce particle-hole symmetry by averaging $G(\tau)$ and $G(\beta-\tau)$. 

\begin{figure}
    \centering
    \includegraphics[width=6.5cm, angle=-90]{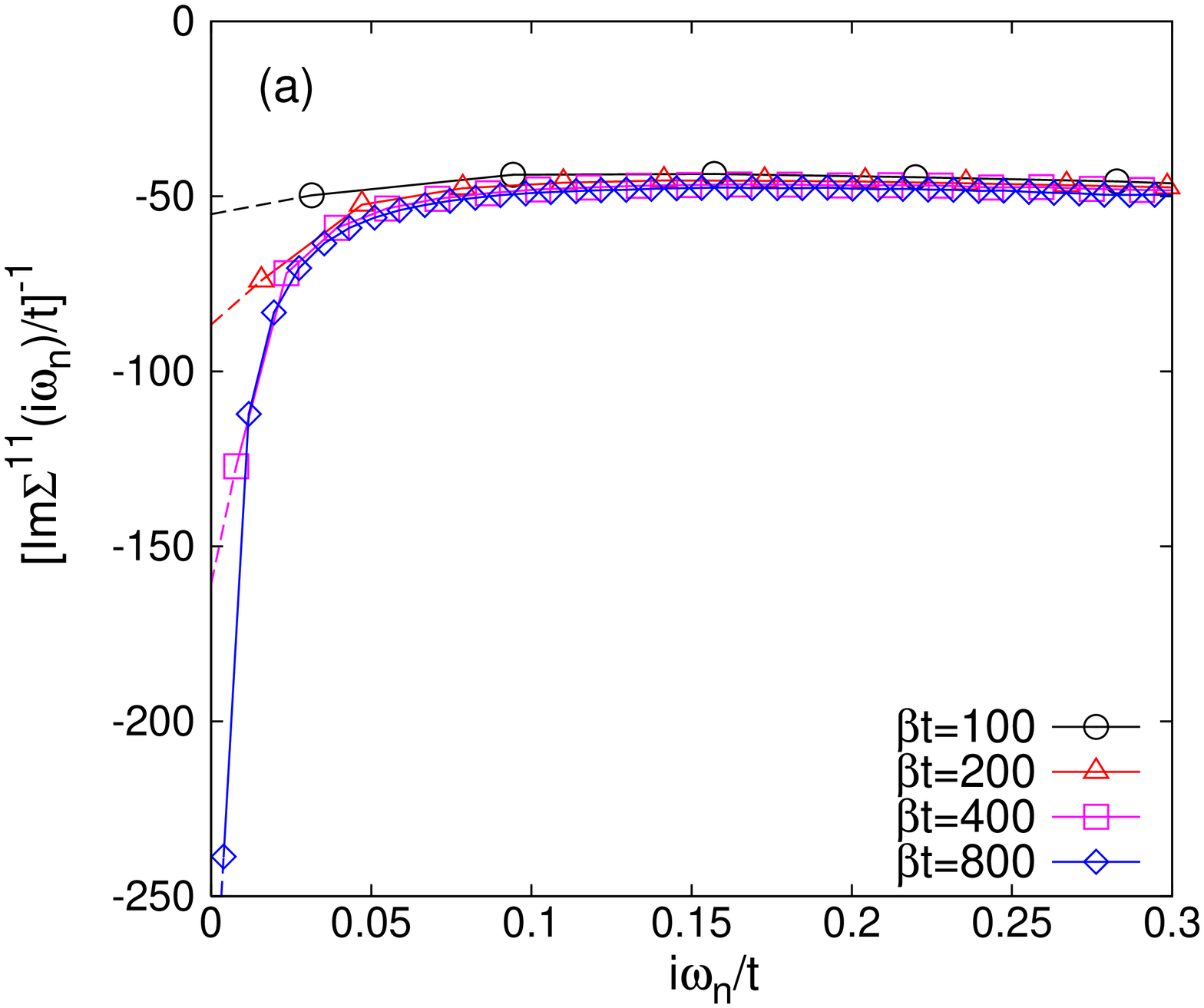}
    \includegraphics[width=6.5cm, angle=-90]{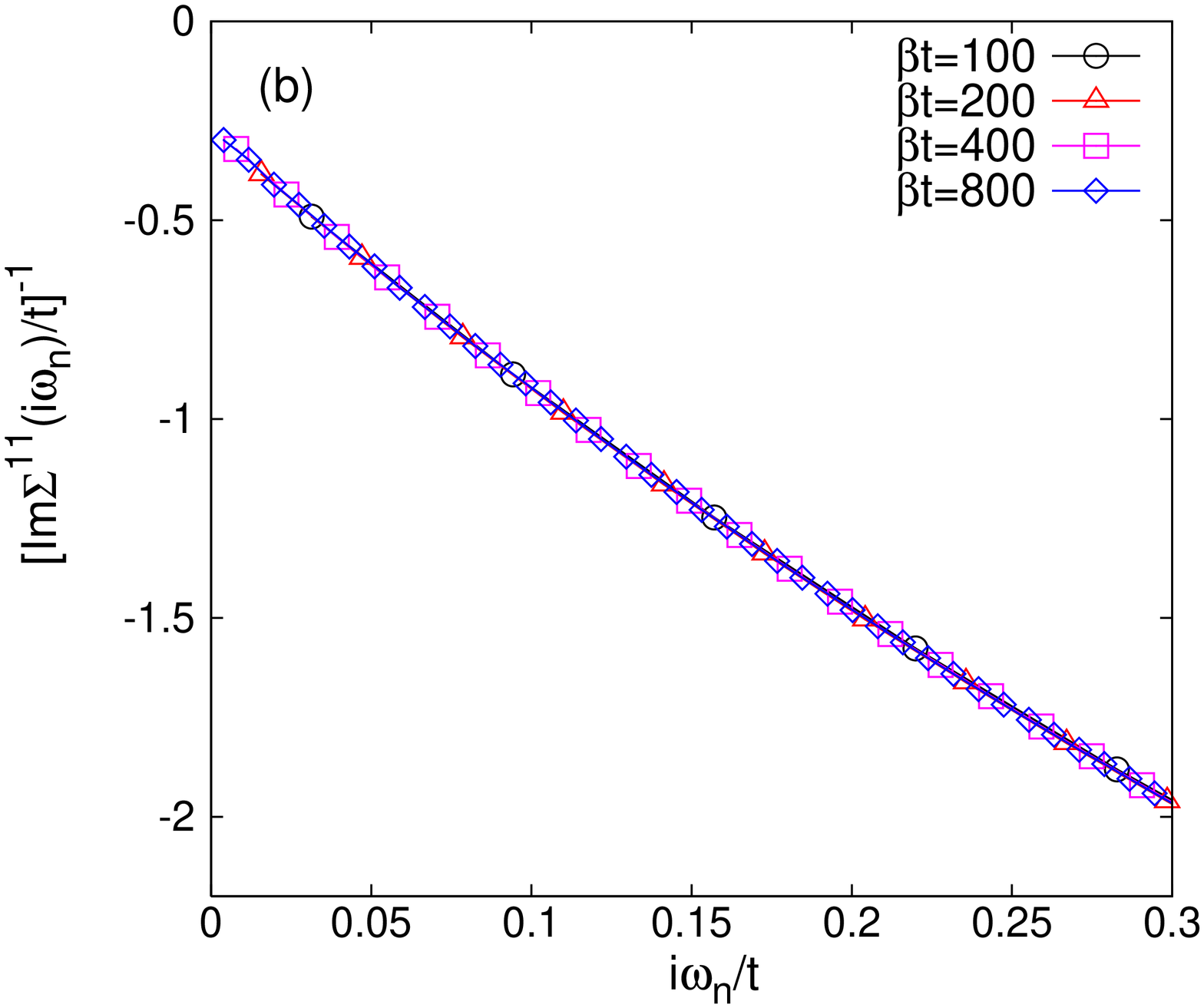}
    \caption{\emph{Reciprocal} of the imaginary part self energy of site 1 calculated for fermions with spin as a function of Matsubara frequencies 
at (a) $U^{12}/t=0.2$ and  (b) $U^{12}/t=0.6$ for temperatures indicated, showing qualitative difference in behavior. The self energy is the same for spin up and down thus only one of them are shown. In panel (a) the dashed lines show an extrapolation to zero frequency. Note the x scale are the same but the y scale are different. Parameters:
$\varepsilon_d=-U^{12}$, $\Gamma^{11}/t=0.04$, $\Gamma^{22}/t=0.25$; }
    \label{sespinful}
\end{figure}

Figure \ref{sespinful} shows the numerical result for the \emph{reciprocal} of the imaginary part self energy of site 1 on Matsubara axis. The self energy of site 2 is very similar and is not shown. Panel (a) is the $U^{12}/t=0.2$ result: we see a clear temperature dependence of the low frequency self-energy.   At the relatively high temperature $\beta t=100$ (open circles, black on-line)  the extrapolated zero frequency limit $[{\rm Im}\Sigma(0)/t]^{-1}$ is  around -50. As temperature is reduced to $\beta t=200$ (triangles, red on-line) $[{\rm Im}\Sigma(0)/t]^{-1}\simeq-87$. Then at $\beta t=400$ (squares, magenta on-line) $[{\rm Im}\Sigma(0)/t]^{-1}\simeq-160$, and at the lowest available temperature $\beta t=800$ (diamonds, blue on-line) $[{\rm Im}\Sigma(0)/t]^{-1}\simeq-300$. It is clear that the absolute value of the intercept at zero frequency increases rapidly as temperature is reduced, consistent with a Fermi liquid picture in which the ground state is a coherent combination of $(0,2)$ and $(2,0)$.

However, turning to Fig.~\ref{sespinful}(b) we see that at the relatively larger interaction strength $U^{12}/t=0.6$ a clearly different behavior occurs. The large difference in y-scales between panels (a) and (b) implies that the self energy is much larger at $U^{12}=0.6$. The traces are approximately linear in frequency and the zero frequency intercept is small and temperature independent.  Fitting the low frequency extrapolation gives $[{\rm Im}\Sigma(i\omega_n)/t]^{-1}\simeq-6.4\omega_n-0.28$, indicating that the $\Sigma\sim i\omega_n$ expected in  a Fermi liquid does not occur   and that  either the ground state behavior has been accessed or at least that any Kondo scale is far below the lowest measurement temperature. The non-vanishing intercept is consistent with the strong scattering expected if the ground state is an incoherent combination of the two valence states. 
\begin{figure}
    \centering
    \includegraphics[width=6.5cm, angle=-90]{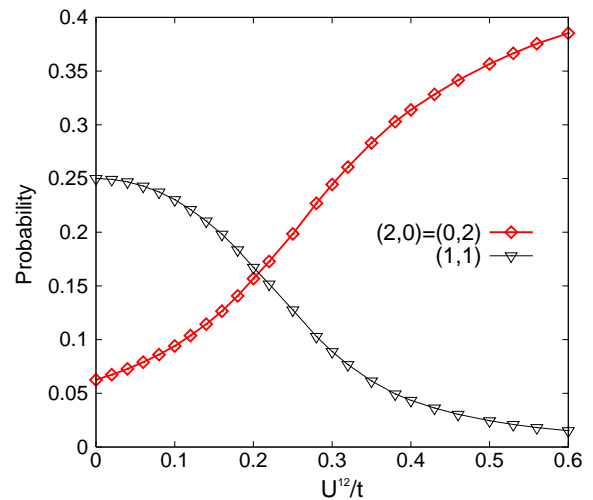}
    \caption{Probability of singlet eigenstates as a function of interaction strength at particle-hole symmetric point, showing dominance of $(0,2)$ and $(2,0)$ pair of states for $U^{12}$ larger than the critical value. Parameters:
$\Gamma^{11}/t=0.04$, $\Gamma^{22}/t=0.25$;
$T=0.01t$. }
    \label{eigenspinful}
\end{figure}

It is natural to associate the onset of non-Fermi-liquid behavior with dominance of the $(0,2)/(2,0)$ pair of states, as we have seen that in the strong coupling limit tunnelling between these states is suppressed. In support of this idea we present in Fig.~\ref{eigenspinful}  the contribution of  each singlet state to the partition function. In the non-interacting case, eigenstate $(1,1)$ has probability of $1/4$ and $(2,0)$ and $(0,2)$ each has probability of $1/16$. As the
interaction is increased, the probablity of $(1,1)$ drops while that of $(2,0)$ and $(0,2)$ increases (and will approach $1/2$ in the infinite interaction limit). Although there can be no sharp transition in occupation probabilities in the impurity models we consider, the  two states $(2,0)$ and $(0,2)$ become increasingly important as $U^{12}$ is increased and are much more important than the $(1,1)$ state in the regimes where non-Fermi-liquid behavior occurs.

\begin{figure}
    \centering
    \includegraphics[width=6.5cm, angle=-90]{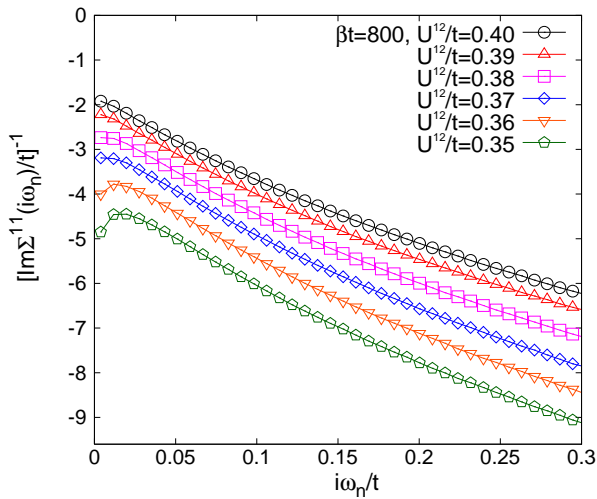} 
    \caption{\emph{Reciprocal} of the imaginary part self energy of site 1 as a   function of Matsubara frequencies for interactions indicated, showing change as critical point is approached. Parameters:
$\varepsilon=-U^{12}$, $\Gamma^{11}/t=0.04$, $\Gamma^{22}/t=0.25$;
$T=t/800$.}
    \label{seUscanspinful}
\end{figure}

Figure~\ref{seUscanspinful} shows the reciprocal of the imaginary part self energy of site one for a finely spaced series of interaction at  the very low temperature $T=t/800$.  The analysis of the spinless fermion case indicated that a weak upturn in a plot of this nature indicated a Kondo temperature as low as $1/5000$ of the measurement temperature.  The data indicate that while for $U^{12}/t=0.35,0.36$ there is still some hint of Fermi-liquid behavior, any Kondo scale drops extremely rapidly and by $U^{12}/t=0.4$ any Kondo temperature is lower  than $1/(800\times5000)\sim3\times10^{-7}t$.

\begin{figure}
    \centering
    \includegraphics[width=6.5cm, angle=-90]{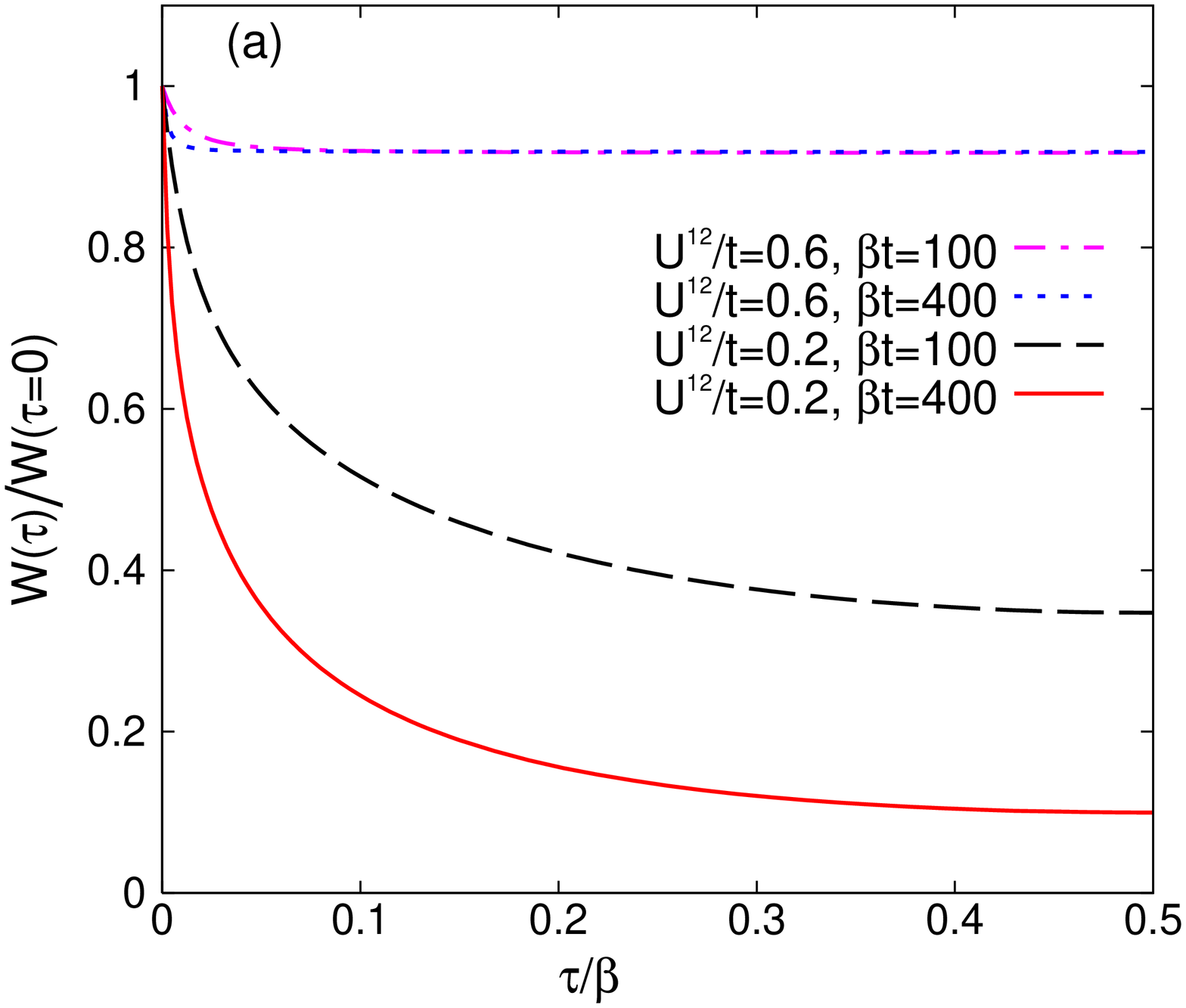}
    \includegraphics[width=6.5cm, angle=-90]{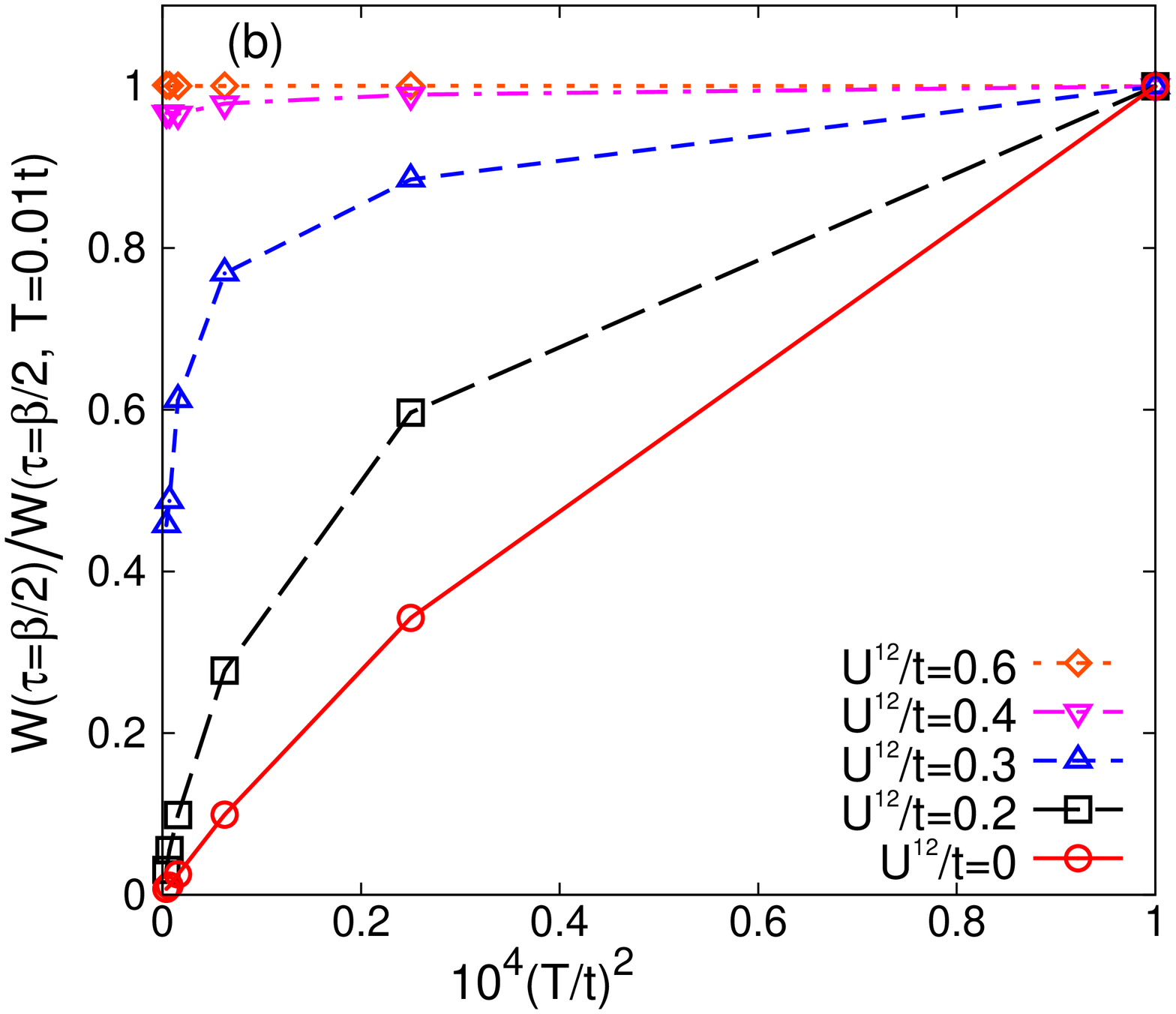}
    \caption{Imaginary time density-density correlation function $W$ (Eq \ref{W}) computed at putative quantum critical point  for  fermions with spin  at parameters indicated, showing weak $T$ and $\tau$ dependence in non-Fermi-liquid phase and strong $T$ and $\tau$ dependence in Fermi liquid phase. (a) Full imaginary time dependence of   $W(\tau)$ normalized to its $\tau=0$ value. (b) Correlation function evaluated at midpoint of imaginary time interval, normalized to value at $T=0.01t$, showing weak $T$ dependence in non-Fermi-liquid phase and $T^2$ dependence at low $T$ in Fermi liquid phase. Parameters $\Gamma^{11}/t=0.04$, $\Gamma^{22}/t=0.25$.}
    \label{corrspinful}
\end{figure}

Figure~\ref{corrspinful} (a) shows the correlation function $W(\tau)$ for fermions with spin . Comparing to Fig.~\ref{corrspinless}(a) we see that while the $U^{12}/t=0.2$ result shown here is rather similar to that of $U/t=0.4$ shown in Fig.~\ref{corrspinless}(a), there are fundamental differences between the $U^{12}/t=0.6$ result shown here and that of $U/t=1.2$ shown in Fig.~\ref{corrspinless}(a). The motivation of comparing $U=2U^{12}$ is that they produce the same Hartree shift, but one should bear in mind that the difference in Hamiltonian will lead to a possible difference in $T_K$, although an evaluation of $T_K$ in the spinful case is lacking in the literature. The flat range in the middle of the $U^{12}/t=0.6$ and its weak temperature dependence, shows strong evidence that fluctuations between the two density eigenstates   are greatly suppressed.

The physics can  more clearly be seen in Fig.~\ref{corrspinful}(b) which shows the temperature dependence of $W(\tau=\beta/2)$  at the particle-hole symmetric point. Fermi liquid theory predicts $W(\tau)\sim1/\tau^2$, thus $W(\beta/2)\sim T^2$ as $T\rightarrow0$, while if the spin fluctuation are frozen, $W(\beta/2)$ becomes a constant at low temperature. The two smallest interactions shown  in Fig.~\ref{corrspinful} ,  $U^{12}/t=0,0.2$ (solid line with circle, red on-line, long dashed line with square, black on-line) clearly reveal the expected  $T^2$ behavior. The next smallest interaction, $U^{12}/t=0.3$ (dashed line with up-pointing triangle, blue on-line), reveals a downturn suggestive of a Fermi liquid ground state, although the $T^2$ regime is not reached. However the stronger interactions $U^{12}/t=0.4,0.6$ (dash-dotted line with down-pointing triangle, magenta on-line and dotted line with diamond, orange on-line) traces are flat down to the lowest accessible temperature $T/t=0.000625$. Combining Fig.~\ref{sespinful} and Fig.~\ref{corrspinful} it is clear that a phase transition happens between $U^{12}/t=0.3$ and $U^{12}/t=0.4$ and the system goes from a  Fermi-liquid like state ($U^{12}<U^{12}_c$) to a ``frozen orbital'' non-Fermi-liquid state ($U^{12}>U^{12}_c$).

\begin{figure}
    \centering
    \includegraphics[width=6.5cm, angle=-90]{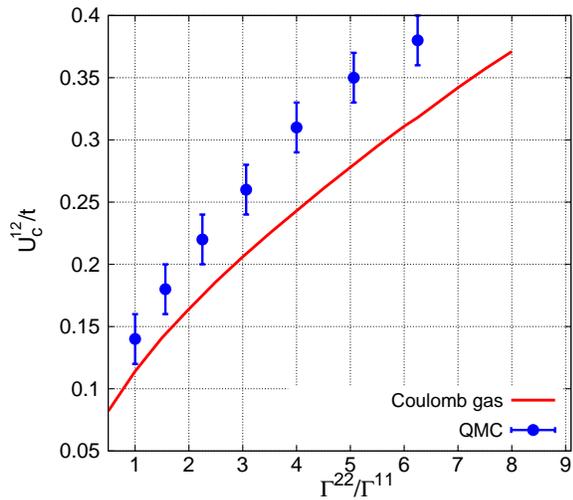}
    \caption{The critical interaction strength $U_c$ as a function of $\Gamma^{22}/\Gamma^{11}$ with $\Gamma^{11}/t=0.04$ fixed. Red line: result from analytical formula. Blue points with error bars: QMC numerical results. Parameters: $\varepsilon=-U^{12}$, $U^{11}=U^{22}=0$.  }
    \label{compAQ}
\end{figure}

The critical interaction strength $U^{12}_c$ is numerically found to be in the range $0.36t<U^{12}_c<0.39t$. This is qualitatively consistent with the analytic estimate  $U_c^{12}=0.32t$. A systematic comparison of the Coulomb gas prediction of $U_c^{12}$ (based on mean field estimates of the phase shifts obtained from the solutions of Eqs.~\eqref{meanfieldspinlessxi} and \eqref{meanfieldspinlesseta}) (note that $2U^{12}$ replaces $U$ in this case) and QMC numerical results as a function of $\Gamma^{22}/\Gamma^{11}$ with $\Gamma^{11}/t=0.04$ fixed is shown in Fig.~\ref{compAQ}. We see a qualitative agreement between the line and the dots; The difference may come from the approximations we made in the mapping to Coulomb gas: We approximated the integration path by a hopping between two $\tau$-independent minima found by minimizing the potential part of the action functional, and the hopping path are assumed to be linear with width $\tau_0$, while in reality the hopping path is more complicated.

\begin{figure}
    \centering
    \includegraphics[width=7cm, angle=-90]{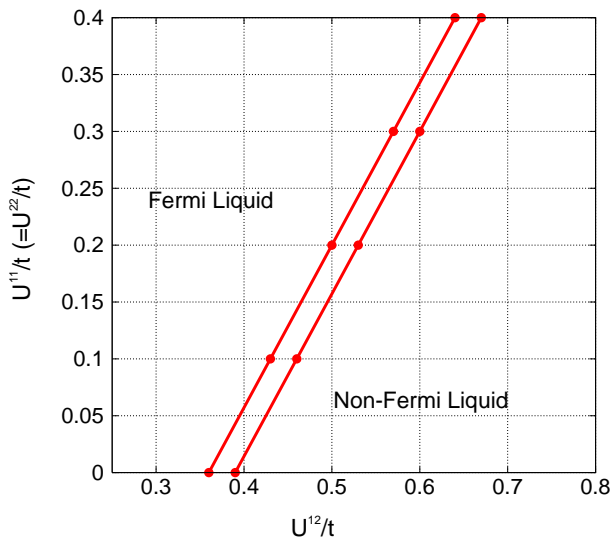}
    \caption{Phase diagram of interaction strength $U^{11}$ versus
    $U^{12}$. $U^{11}=U^{22}$ has been assumed. $\varepsilon^{ii}=-U^{12}-U^{ii}/2$, $(i=1,2)$. Note the offset of zero of x-axis. In the non-Fermi-liquid regime a quantum critical point occurs as level energies are changed. Parameters: $\Gamma^{11}/t=0.04$, $\Gamma^{22}/t=0.25$.}
    \label{phasediag}
\end{figure}

Fig.~\ref{phasediag} shows a phase diagram indicating the locus of criticality in the plane of on-site interaction strength $U^{11}$ and intersite interaction strength  $U^{12}$ at the particle-hole symmetric point $\varepsilon^{ii}=-U^{12}-U^{ii}/2$, $(i=1,2)$. $U^{11}=U^{22}$ has been assumed for simplicity. Above and to the left of the two lines the physics is Fermi-liquid like; below and to the right a quantum critical point occurs as level energies are changed. The regime between the two lines is a crossover region in which  we cannot determine whether it is Fermi liquid within the temperatures studied ($T/t\geq1/800$). We can see that as $U^{11}$ increases, the critical $U^{12}$ needed for a non-Fermi-liquid behaviour increases.

\section{Conclusions \label{Conclusion}}

In this paper we have studied the possibility of ``quantum criticality'', defined here as a sharp transition in level occupancy as the parameters of a multiorbital quantum dot are varied. A sharp transition requires a multistability, with more than one locally stable solution, and thus in particular requires that a symmetry or physical mechanism prevents tunnelling between the different potential solutions. In the situation of relevance here, the physical mechanism is the orthogonality effect arising from dissipative coupling to leads. Our work was motivated by an interesting proposal arising in the context of a two-dot interferometer,\cite{Golosov06} but we observe that multistability is of broader interest in the context of potential molecular devices.  We presented an analysis of the types of level-occupancy-related quantum critical points that could arise and focussed on the particular situation introduced by Ref.~\onlinecite{Golosov06}. We used analytical arguments based on a mapping to a Coulomb gas, as well as numerical calculations to show that  while the originally studied case of spinless fermions exhibited only a Fermi liquid behavior, a model of  fermions with spin could in appropriate circumstances exhibit a sharp switching. The result for spinless electrons is in agreement with previous numerical renormalization group studies  by Refs.~\onlinecite{Sindel05, Meden06,Karrasch07a,Karrasch07b}. (Note that the jumps indicated in Ref.~\onlinecite{Sindel05} arise from a multiple-lead multiple-level situation in which the width of one level accidentally vanishes.) The key requirements are an intersite Coulomb interaction which is larger than the on-site one and the absence of ``pair hopping'' terms in the dot Hamiltonian. This leads to an ``orthogonality'' exponent greater than the critical value of two and hence to localization in macroscopic quantum tunneling sence. These essential ingedients may be difficult to realize in practice, although a  strong local electron-phonon coupling could lead to a polaronic suppression of the on-site Coulomb interaction; thus a system that incorporated this physics might be an appropriate realization.  An interesting feature of our numerics is that signatures of a Kondo effect are visible in the fermion self energy at temperatures orders of magnitude above the Kondo scale.

Recently we became aware of a preprint \cite{Goldstein09b} reporting also that the spinless fermion model does not exhibit quantum criticality, and proposing that quantum criticality could also be realized in a three-lead model (which would similarly increase the orthogonality exponent).

{\it Acknowledgements:}  We thank P. Werner, J. Lin and C. Lin for helpful discussions. This work is supported by NSF Grants No. DMR-0705847 and No. CHE-0641523 and the New York State Office
of Science, Technology and Academic Research (NYSTAR).

\end{document}